\documentclass{winnower}

\usepackage{graphicx}

\usepackage{amssymb}
\usepackage{natbib}
\usepackage{leftidx}
\usepackage{amsmath}
\usepackage[usenames, dvipsnames]{color}
\usepackage{soul} 

\newcommand{\x}{\mathbf{x}}

\newcommand{\median}{\text{median}}
 \newcommand{\mad}{\text{mad}}

 \newcommand{\mean}{\text{mean}}
 \newcommand{\std}{\text{std}}

\usepackage[printwatermark]{xwatermark}
\usepackage{xcolor}
\usepackage{graphicx}
\usepackage{lipsum}
\usepackage{multirow}

\usepackage[normalem]{ulem}

\usepackage[colorinlistoftodos]{todonotes}

\begin{document}

\title{Clustering of imbalanced high-dimensional media data}

\author{Sarka Brodinova}
\affil{Institute of Statistics and Mathematical Methods in Economics, Vienna University of Technology, Austria}
\author{Maia Zaharieva}
\affil{Institute of Statistics and Mathematical Methods in Economics, Vienna University of Technology \&  Multimedia Information Systems Group, University of Vienna,  Austria}
\author{Peter Filzmoser}
\affil{Institute of Statistics and Mathematical Methods in Economics, Vienna University of Technology, Austria}
\author{Thomas Ortner}
\affil{Institute of Statistics and Mathematical Methods in Economics, Vienna University of Technology, Austria}
\author{Christian Breiteneder}
\affil{Institute of Statistics and Mathematical Methods in Economics, Vienna University of Technology, Austria}

\date{}

\maketitle

\begin{abstract}
\sloppy Media content in large repositories usually exhibits multiple groups of strongly varying sizes. Media of potential interest often form notably smaller groups. Such media groups differ so much from the remaining data that it may be worthy to look at them in more detail. In contrast, media with popular content appear in larger groups.
Identifying groups of varying sizes is addressed by clustering of imbalanced data. Clustering highly imbalanced media groups
is additionally challenged by the high dimensionality of the underlying features. In this paper, we present the Imbalanced Clustering (IClust) algorithm designed to reveal group structures in high-dimensional media data. IClust  employs an existing clustering method in order to find an initial set of a large number of potentially highly pure clusters which are then successively merged. The main advantage of IClust is that the number of clusters does not have to be pre-specified and that
no specific assumptions about the cluster or data characteristics
need to be made. Experiments on real-world media data  demonstrate that in comparison to existing methods, IClust is able to better identify media groups, especially groups of small sizes.
\end{abstract}

\section{Introduction}\label{sec:introduction}

Nowadays, large media repositories,  such as YouTube\footnote{http://www.youtube.com} and Vimeo\footnote{http://www.vimeo.com}, are facing continuous additions of new, unknown material. Media, such as videos, sounds, and images,  are straightforward to capture and share due to recent developments and advances of existing technologies. The processing and analysis of large amounts of unknown material is very challenging  when trying to understand media content and 
when there is no prior knowledge available.  Additionally, media of potential interest can easily get lost in the mass of available data. In this context, interestingness is a data-driven concept describing content which is that much different from the remaining data and, therefore, it is potentially worthy to look at in more detail. 
As any other data collection, media data commonly exhibits multiple groups.
The underlying groups of media have strongly varying sizes
and, therefore, such a data set is considered as highly imbalanced.
While larger groups represent a common type of media
(e.g. video recordings of a popular music band),
very small groups (even a size of one) commonly indicate atypical content and, thus, potential interesting material (e.g. video recordings of a non-famous street musician). The task of identifying small groups is additionally challenged by the characteristics of multimedia content. Media data are commonly represented by means of high-dimensional features. Conventional clustering methods based on traditional model assumptions
usually fail in such a situation~\citep{DBLP:journals/tkdd/KriegelKZ09}. 
The focus of this paper is on developing a cluster algorithm
which addresses two core challenges: 1)~clustering in a high-dimensional data space and 2)~clustering imbalanced data with special attention on mining small groups.


Detecting clusters in high-dimensional space is commonly addressed by subspace or projected clustering algorithms which search for clusters in a subset of dimensions. Therefore, such methods are suitable for high-dimensional data where there is a large proportion of noise variables.  However, these methods usually require for a parameter specification~\citep{DBLP:journals/sigkdd/ParsonsHL04} which may be problematic, especially for media data where no prior knowledge is available. On the contrary, if the number of noise variables is not too high, model-based clustering methods \citep[e.g.][]{Fraley00model-basedclustering} or density-based algorithms~\citep{DBLP:journals/widm/KriegelKSZ11}, could still achieve promising results. Nevertheless,  density-based approaches might be more appropriate for media data. Such methods commonly do not rely on any prior knowledge (e.g. number of clusters, shapes of clusters, and distribution of clustered points), which might be very beneficial when clustering media data.

Clustering imbalanced data, where group sizes are
very different, causes additional challenges.  Even though the research area of imbalanced 
clustering is not recent, there are still open issues which need to be addressed 
in the development of new methods~\citep{Krawczyk2016}.  A very first problem addressed by \citet{Krawczyk2016}, which usually occurs in centroid-based methods, is the so-called uniform effect. This means that a clustering algorithm generates clusters of similar sizes. Some observations from larger groups are mixed with those from smaller groups. 
In order to prevent the effect, \citet{Krawczyk2016} 
proposed a hybridization of centroid-based and density-based methods.
Another proposal can be found in literature \citep[e.g.][]{DBLP:conf/icarcv/WangC14,DBLP:conf/icassp/QianS14}. However, most approaches assume prior knowledge of the number of clusters in order to handle varying levels of imbalanced data. 
Finally, \citet{Krawczyk2016} pointed out on the potential of discovering very small groups which could be useful for further analysis. Indeed, media collection is a good example of imbalanced data where small groups are of potential interest.


In this paper, we propose the Imbalanced Clustering (IClust) algorithm, an approach which 
is able to identify groups of potentially strongly varying sizes.
The procedure first employs an existing method which is forced to produce
a large initial set of potentially pure clusters. Subsequently, clusters are successively merged using two merging conditions based on the outlier detection method -  Local Outlier Factor~\citep{DBLP:conf/sigmod/BreunigKNS00}. The algorithm stops when two merging conditions are not satisfied and, therefore, the final number of clusters does not need to be pre-specified. This is an advantage over
most existing methods. Moreover,  the proposed approach detects clusters of strongly varying sizes without any specific assumptions about the cluster or data characteristics,
which is very important for clustering media data.

The remainder of this paper is organized as follows. 
Section~\ref{sec:approach} motivates the design of the proposed algorithm and describes it in detail.  
In Section~\ref{sec:parameters}, we select optimal parameters for the proposed algorithm, and results on real-world media data sets are presented in Section~\ref{sec:comparison}.
Section~\ref{sec:discussion_conclusion} concludes the paper.

\section{Proposed clustering algorithm}\label{sec:approach}

The idea for our algorithm originates in the need to efficiently detect small groups in media data, containing potentially interesting information.
In order to identify highly imbalanced groups, we employ  the Local Outlier Factor~\citep{DBLP:conf/sigmod/BreunigKNS00}
originally designed to reveal outliers deviating from clusters.

\subsection{Background on Local Outlier Factor (LOF)}
Preliminary experiments indicated that LOF is a highly effective approach for the identification of very small  groups in media data as outliers in comparison to other existing approaches. In general, LOF determines the degree of outlyingness of an observation. The degree reflects to which extent an observation is isolated from its predefined number of the nearest observations. 
Let~$\mathbf{X}=(\x_1, \dots, \x_n)^\top  \in \mathbb{R}^p $ be  a data set  of $n$ observations from the Euclidean space of $p$ dimensions. The LOF score for each observation $\x_i  , i=1,\dots,n$, is defined according to \citet{DBLP:conf/sigmod/BreunigKNS00} as:
\begin{equation}
\label{eq:Lof}
LOF_{q}(\mathbf{x}_i)= \frac{1}{N_{q}(\x_i)} \sum_{\x \in N_{q}(\x_i)}\frac{lrd_{q}(\x)}{lrd_{q}(\x_i)},
\end{equation}
where $N_{q}(\x_i)$ denotes the local neighborhood for $\x_i$ defined by its $q$ nearest observations, i.e. neighbors, and $lrd_{q}(\x_i)$ corresponds to so-called local (reachability) density of $\x_i$. The local density of $\x_i$ reflects how distant $\x_i$ is with respect to its $q$ nearest neighbors on average, taking into account the distances of its neighbors to their nearest observations, see ~\cite{DBLP:conf/sigmod/BreunigKNS00} for more details.
In~general, if  $\x_i$ belongs to a cluster, i.e.  $\x_i$ is surrounded by or close enough to its neighbors $\x \in N_{q}(\x_i)$, the local density of  $\x_i$ is similar to the local densities of its neighbors. As a consequence, $LOF_{q}(\mathbf{x}_i)$ achieves a value of approximately 1. In contrast, an observation  $\x_i$ deviating from a cluster has considerably
different local densities than its neighbors since $\x_i$ is highly isolated from its neighbors. Therefore, such an observation receives $LOF_{q}(\mathbf{x}_i) >> 1$ and can thus be declared as outlier.

In addition to the ability of  LOF to detect outliers based on large LOF scores,
the LOF approach exhibits several properties which might be very useful in clustering imbalanced data. First, LOF 
is suitable for the contaminated data with the clusters of varying sizes and densities \citep{Hasan2009994}. Second, the decision of declaring an observations as an outlier seems to be insensitive 
to the choice of 
the predefined number of neighbors $q$ necessary for calculating LOF scores \citep{Hasan2009994}. Next, LOF does not rely on any specific 
assumptions on the cluster characteristics~\citep{DBLP:conf/sigmod/BreunigKNS00},
which is particularly important for high-dimensional media data where such assumptions
could not be verified. Finally, \citet{SAM:SAM11161} demonstrated that 
LOF achieves promising results if the number of informative variables
in the high-dimensional data is not too low. We expect that media data also
contain many informative variables, while, for example, for gene expression data
this might not necessarily be the case.

Despite the mentioned advantages of LOF and the fact that the method is capable of detecting very small 
groups in media data as outliers, we still need to adjust the usage of LOF in order to recover a whole group structure in imbalanced media data.

\subsection{Naive approaches}
\begin{figure}[h]
\centering
  \includegraphics[width=0.89\textwidth]{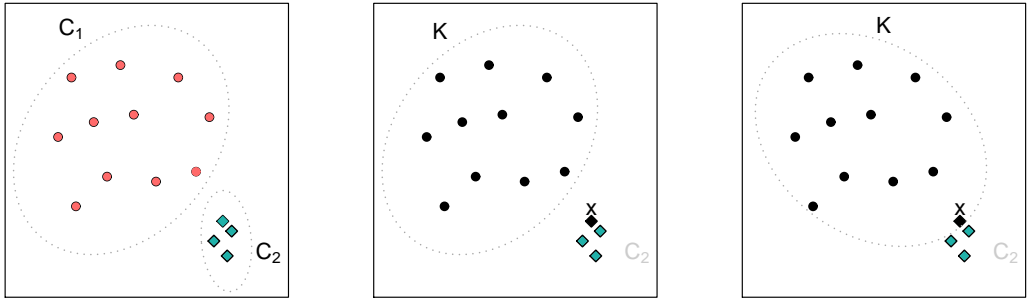}
\caption{
Example of a data set with two groups of different sizes and densities, $C_1$ and $C_2$ (left); the correctly identified group $C_1$ as a cluster $K$ with its next closest point $\x$ from group $C_2$ displayed in black color (middle); a wrong assignment of $\x$ to $K$ due to a LOF-based decision (right)}\label{fig:recursive}
\end{figure}

A naive idea would be to remove the detected outliers from the data, and to use existing cluster methods for clustering the larger groups. However, this leads to difficulties since the resulting group sizes might still be very different and because many clustering algorithms assume a certain shape of the clusters. Another idea would be a recursive identification of the clusters,
by starting to build the first cluster in the most dense and compact region. In the following, LOF can be used to decide which points are still members of this cluster, and which point is too far away (outlier) in order to form a new cluster.  However, this decision can become unreliable as illustrated in Figure~\ref{fig:recursive}. Suppose that there are two groups $C_1$ and $C_2$ with different sizes and densities (left picture). Assume that the first cluster $K$ has been constructed and it needs to be decided whether or not  observation $\x$ still belongs to cluster $K$ (middle picture). If the neighborhood size used to compute the LOF score is not small enough, the point $\x$ would be assigned to cluster $K$ because of the different point densities of the underlying clusters (right picture). Using a very small neighborhood size instead would again be unreliable because the decision would be based on too little
information. Therefore, for being able to identify groups of very small sizes, even of size one, we need to modify the concept, which leads to the proposed IClust approach.

\subsection{The IClust algorithm}

The proposed algorithm is conducted in two steps. In the first step of IClust, we identify a large number of initial clusters by employing an existing clustering method.
We suggest to take a simple existing method which allows to control the number of clusters, e.g. $k$-means.
The large number of initial clusters leads to potentially highly pure regions of comparable densities
 of clustered points.  In the second step, we merge clusters by employing the LOF approach.  
In each merging step it is investigated if two closest clusters share the same local 
densities; if this is the case, the clusters are merged. This avoids the wrong assignment
as indicated in Figure~\ref{fig:recursive}. The algorithm iteratively tries to merge the 
next two closest clusters until there are no clusters to be merged. We give a detailed description of each step in the following sections.

\subsubsection{Identifying the initial set of clusters} \label{subsub:init_method}
In the first step, the data set is split into an initial set of clusters by applying an existing clustering algorithm  
to subdivide the underlying data set into a large number of clusters.  Although such a partitioning leads to over-clustering of the data, it allows for the detection of (highly) pure clusters. We propose to use such a clustering algorithm that 
is less computationally demanding and requires 
for the number of clusters only in order to enable to control the number of initial clusters and not to be influenced by a wrong choice of parameters which is usually data-dependent,
e.g. $k$-means.

The number of initial clusters $k_{init}$ needs  to be set large enough (larger than the true underlying number of clusters) in order to increase the probability of obtaining highly pure clusters. However, the value of $k_{init}$ should not be too large to have a sufficient number of observations, i.e. information, in most clusters for 
the merging procedure.
There are several possibilities for the determination of the number of initial clusters. For example, \citet{DBLP:conf/icvs/BloisiI08} suggest to partition the data into $n/4$ clusters. In Section~\ref{sec:parameters} we investigate different selection strategies for $k_{init}$ and their influence on the clustering solution.

We will experiment with several well-known clustering methods to identify the optimal choice for a starting clustering algorithm in Section~\ref{sec:parameters}. We consider two partitioning methods, $k$-\emph{means}~\citep{Hartigan/Wong:79}\footnote{$k$-means is implemented in R package \texttt{stats}~\citep{stats}.} and \emph{Partitioning Around Medoids}~\citep{DBLP:books/wi/KaufmanR90}\footnote{PAM is implemented in the R package \texttt{cluster}~\citep{cluster}.}, two hierarchical methods, \emph{complete linkage} and \emph{Ward's method}~\citep{Murtagh2014}\footnote{Complete linkage and Ward's method are implemented in the R package \texttt{cluster}~\citep{cluster}.}, and the model-based clustering method \emph{Mclust}. All methods have certain drawbacks in terms of generating specifically shaped clusters and they suffer from the so-called uniform effect. 
By incorporating these methods in the first step we can enhance their performance in 
a highly imbalanced scenario. A large number of clusters generated by these methods avoids the uniform effect.
Furthermore, over-clustering can prevent from being affected by the assumption about the shapes of clusters.

\begin{figure}[h]
\centering
  \includegraphics[width=0.89\textwidth]{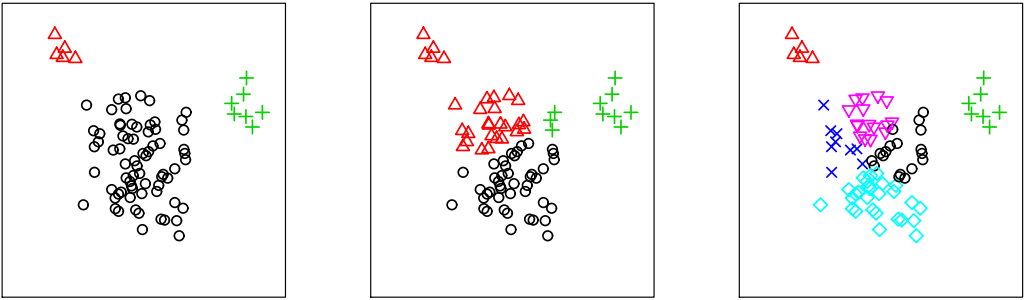}
\caption{The effect of $k$-means applied on a 2D imbalanced data set. Imbalanced data set with three groups (left), $k$-means with three clusters (middle) and $k$-means with six clusters (right)}\label{fig:effect}
\end{figure}

To illustrate both aspects, we apply $k$-means on an imbalanced data set. We consider a simple 2D data set with three groups of different sizes as shown in Figure~\ref{fig:effect} (left). Applying $k$-means with the true number of clusters, i.e. $k=3$, results in the wrong assignment of observations from the large group to the smaller groups, see Figure~\ref{fig:effect} (middle), because $k$-means 
tends to produce spherically shaped clusters. In contrast, $k$-means with a larger number of clusters, e.g. $k=6$, results in a solution as shown in  
Figure~\ref{fig:effect} (right), where
the smaller groups are correctly detected and the large group is split into four small but pure clusters. Therefore, in the next step, we aim at merging small clusters that likely belong to the same group while keeping well-isolated and, thus, potentially correctly detected small groups.

\subsubsection{Merging procedure}\label{subsec:merging}

The aim of the second step is to iteratively merge clusters that are close to each other and share the same local densities. The underlying assumption is that such clusters contain observations from the same group. 
In order to consider both distances and local densities, we propose to investigate if a point from one cluster can be considered as part of a second cluster and vice versa by employing the LOF. The purpose of investigating two clusters twice is to avoid that a cluster of low density is merged with a cluster of high density, as discussed at the beginning of this section.  
Therefore, we introduce the two merging conditions which need to be satisfied to merge the two closest clusters.

Let $\{K_j | j=1,\dots,k_{init}\}$ be the initial set of clusters, where $K_j=\{ \x_{i_j} | i_j \in  I_j \}$ contains observations  from the index set $I_j=\{ 1_j,2_j,\dots,{|K_j|}_j\}$. The merging procedure starts by finding the two closest clusters, $K_l$ and $K_m$, based on the minimum distance between each pair of observations coming from different clusters (single linkage approach). In addition, the two closest points, $\x_o \in K_l$ and $\x_p \in K_m$, are identified such that
\begin{equation}
d( \mathbf{x}_o,\mathbf{x}_p )= \smash{\displaystyle\;  \min_{\x_{i_l} \in K_l, \x_{i_m} \in K_m}} \;  d(\mathbf{x}_{i_l},\mathbf{x}_{i_m}).
\end{equation}
Subsequently, we investigate whether or not the two clusters should be merged. 

\begin{figure}[h]
\centering
  \includegraphics[width=0.89\textwidth]{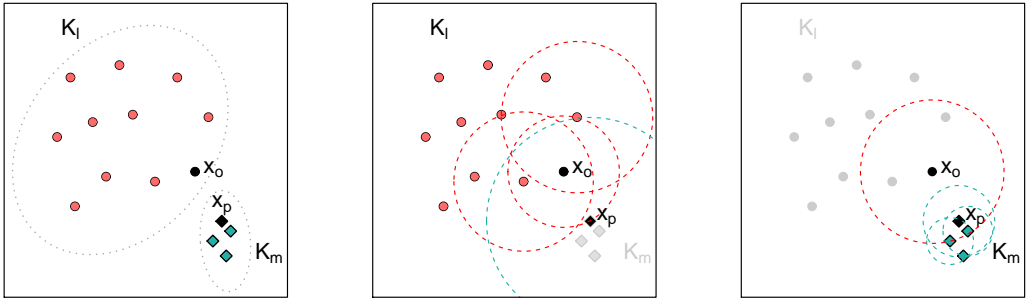}
\caption{
Illustration of the proposed merging procedure for two clusters $K_l$ and $K_m$ with the corresponding closest points $\x_o \in K_l$, $\x_p \in K_m$ (left). The neighborhood of $\x_p$ with its three nearest neighbors is displayed as circles. $\x_p$ is close to its three neighbors (from $K_l$) located in the circle around $\x_p$ (middle).   $\x_o$ is highly isolated from its three neighbors (from $K_m$) located in the circle around $\x_o$ (right) 
}\label{fig:merging}
\end{figure}

For illustration, we consider the simple example in Figure~\ref{fig:merging}~(left) showing two clusters $K_l$ and $K_m$ with the corresponding closest points $\x_o \in K_l$ and $\x_p \in K_m$.  
Figure~\ref{fig:merging} (middle) shows the idea of investigating whether or not $\x_p$ can be part of $K_l$ considering that the neighborhood is defined by three closest neighbors, denoted as $q=3$. The plot particularly indicates that $\x_p$ is close to its three neighbors from $K_l$, which are located in the circle around $\x_p$. In addition, it seems that the observations located in the neighborhood (displayed as circles) form a compact region of similar local densities. As a result, the LOF score of $\x_p$ should be approximately 1. In such a case, we conclude that $\x_p$ can be considered as part of the second cluster $K_l$.

%

Formally, we calculate the LOF score for observations from $K_l$ and the point $\x_p$
according to Eq. (\ref{eq:Lof}) for a predefined range of the number of nearest neighbors $q$, determined by the maximal number of nearest neighbors $q_{max}$:
\begin{equation}
LOF_{q}(\mathbf{x}_{i}), \quad i\in \{ I_l \cup p \},  \quad q=1,2,\dots,q_{max},
\end{equation}
This results in several values of the LOF score for each observation depending on the range of $q$. The reason for considering different choices of $q$ is to obtain more information about the local densities with respect to various sizes of the neighborhood. Subsequently, we calculate a representative value, $lof(\mathbf{x}_{i})$: 
\begin{equation}
lof(\mathbf{x}_{i})=\frac{1}{|q|}  \sum_q LOF_{q}(\mathbf{x}_{i}) \quad  i\in \{ I_l \cup p \}, \quad q=1,2,\dots,q_{max},
\end{equation}
where $|q|$ denotes the number of different choices of $q$. 
We provide an empirical study on the proper choice of $q$ in Section~\ref{sec:parameters}. For now, suppose that $q$ is given. The value of $lof(\x_{i})$ describes the average similarity between the local density of an observation $\x_{i}$ and the local densities of its neighbors. In addition, $lof(\x_{i})$ has  similar properties as the LOF score since it is a linear combination of the original scores. The higher the value of $lof(\x_p)$, the more different are the local densities of $\x_p$ with respect to neighbors from $K_l$,
and the more likely $\x_p$ is an outlier with respect to $K_l$, i.e. $\x_p$ cannot be a part of $K_l$.
In order to decide if the compared local densities are similar, i.e. $\x_p$ can be  part of $K_l$,
it is necessary to determine how large $lof(\x_p)$ still can be to consider  $\x_p$ as part of $K_l$ . The most convenient option would be to decide on the basis 
of the resulting LOF scores. Therefore, we estimate a critical value $cv^p$ from the values of $LOF_{q}(\x_{i})$, where $i\in \{ I_l \cup p \}$ and $q=1,2,\dots,q_{max}$. There are several possibilities for the determination of the critical value, such as using the arithmetic mean and standard deviation or robust versions thereof. 
The optimal strategy for the estimation of the critical value is presented in Section~\ref{sec:parameters}.
For now, we assume that $cv^p$ is given and we test if \textit{the first merging condition}, $lof(\x_p) < cv^p$, holds. If the first condition is fulfilled, we consider the compared local densities similar and apply the same comparison on $\x_o$ with respect to $K_m$, i.e. we investigate if $\x_o$ can be considered as part of the opposite cluster $K_m$. 

Figure~\ref{fig:merging}~(right) shows that $\x_o$ is considerably isolated from its three closest neighbors from $K_m$ and that the observations do not build any compact region. In such a case we can conclude that there are huge differences in the local densities. Therefore, $LOF_{q}(\mathbf{x}_{o})\gg1$ which
indicates that $\x_o$ can not be considered as a part of the second cluster $K_m$. Formally, we calculate the LOF score for the observations from $K_m$ and the point $\x_o$ for the predefined range $q=1,2,\dots,q_{max}$. The critical value $cv^o$ is estimated in the same way as $cv^p$. Subsequently,  we test if \textit{the second merging condition}, $lof(\x_o) < cv^o$, is satisfied. If this condition is not fulfilled, the two clusters, $K_m$ and $K_l$, are not merged and the next two closest clusters are investigated. The merging procedure stops if the two conditions are not satisfied for any pair of clusters.

\begin{figure}[h]
\centering
  \includegraphics[width=0.89\textwidth]{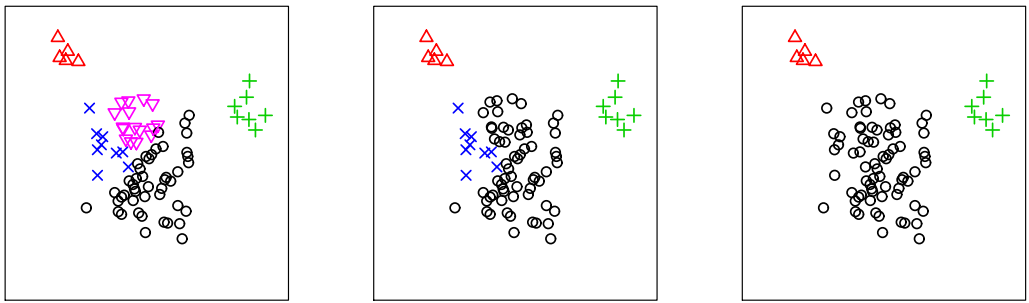}
  \caption{The merging procedure successively merges clusters that are close to each other and share the same distribution. The final solution indicates that the detected clusters correspond to the actual group structure }\label{fig:Iclust}
\end{figure}

The proposed IClust algorithm exhibits several advantages. 
First, the number of final clusters does not need to be pre-specified 
due to the employed merging procedure. Second, the proposed procedure makes no assumptions about the cluster characteristics. The local densities are estimated in a non-parametric way following the definition of LOF. Finally, IClust detects clusters of partly  strongly varying sizes. By using an existing clustering algorithm with  a large number of clusters, we avoid the so-called uniform effect. For illustration, we consider the 2D imbalanced data set and apply $k$-means with $k=6$ to generate an initial set of clusters as in our first example in Figure~\ref{fig:effect}~(right). Figure~\ref{fig:Iclust} shows each merging of the two next closest clusters. The final result indicates that both smaller and larger groups are correctly detected.

\section{Selection of parameters}\label{sec:parameters}
In this section, we investigate different strategies for the parameter selection. The IClust algorithm requires for four input parameters: 1) the critical value, $cv^o$ ($cv^p$), 2) the range of the nearest neighbors, $q$, 3) the number of initial clusters, $k_{init}$, and 4) the starting clustering algorithm. While the first two parameters are employed in the merging procedure, the last two parameters  are used to partition the data set into an initial set of clusters. Since two parameters depend on each other, we always fix one parameter to investigate the second one and vice versa. The optimal parameter setting is chosen based on thorough empirical experiments employing the audio data set. 

The audio data set consists of 4780 observations which is a collection of 12 different audio sounds. Each observation is represented by a feature vector of 679 dimensions. The extracted features capture a wide range of audio properties and operate in the temporal and frequency domains, i.e. the data include features such as zero 
crossings, amplitude, brightness, features from the MPEG7 standard, perceptional features, and various cepstral coefficients. In our experiments we randomly sample observations from the original groups to create imbalanced data sets. The variables of each constructed data set are normalized to mean  $0$ and standard deviation $1$.

\subsection{Critical value}\label{subsub:cv}

The first experiment investigates different strategies for the estimation of the critical values, $cv^o$ and $cv^p$, employed in the merging procedure. The critical values determine whether or not two clusters should be merged.  Both critical values are estimated in the same way, therefore, let $cv$ be a general estimation of the critical value. The value of $cv$ is supposed to be automatically derived from the LOF scores calculated for the observations from two clusters. We consider several possibilities for the estimation of $cv$ including arithmetic mean, \textit{$\mean()$}, empirical standard deviation, \textit{$\std()$}, and robust versions, such as the median, \textit{$\median()$} and the median absolute deviation, \textit{$\mad()$}:
\begin{align}
  cv_1 &=  \underset{q,i} \median \; ( LOF_q(\mathbf{x}_i)) + 2\; \underset{q,i} \mad\; ( LOF_{q}(\mathbf{x}_i))  \label{eg:cv1}\\
 cv_2 &=\underset{q,i}  \mean \; ( LOF_q(\mathbf{x}_i)) + 2\; \underset{q,i}  \std \;  (LOF_{q}(\mathbf{x}_i))\\
 cv_3 &=\underset{i}\median \;  (lof(\mathbf{x}_i))+ 2\; \underset{i} \mad\;  (lof(\mathbf{x}_i))\\
 cv_4 &=\underset{i}\mean \;  (lof(\mathbf{x}_i)) + 2\; \underset{i}   \std\;( lof(\mathbf{x}_i)),
\end{align}
where $i$ is either from the index set $\{ I_l \cup p \}$  for the first merging condition or  from $ \{ I_m \cup o \}$ for the second merging condition. The range for the number of nearest neighbors determined by $q_{max}$ is fixed to $min(|\{ I_l \cup p \}|-1,5)$ and $min(|\{ I_m \cup o \}|-1,5)$, respectively.

Since $cv$ determines whether or not two clusters should be merged, we investigate two situations.  We first simulate the situation when two clusters should not be merged, i.e. the underlying observations are from two different groups. The second situation considers two clusters containing observations from the same group and, therefore, the two clusters are supposed to be merged. For this experiment we employ observations from the audio data. We randomly sample two clusters either from two different audio groups or from the same audio group. The sampled clusters are of varying sizes of $30, 25, 20, 15, 10, 5, 3, 1$. We investigate each possible pairwise combination thereof and perform $10$ replications for each combination. The percentage of correct decisions is considered as a performance indicator for the different strategies for the estimation of the critical value, $cv$. 

\begin{figure}[h]
  \begin{center}
    \includegraphics[width=0.65\textwidth]{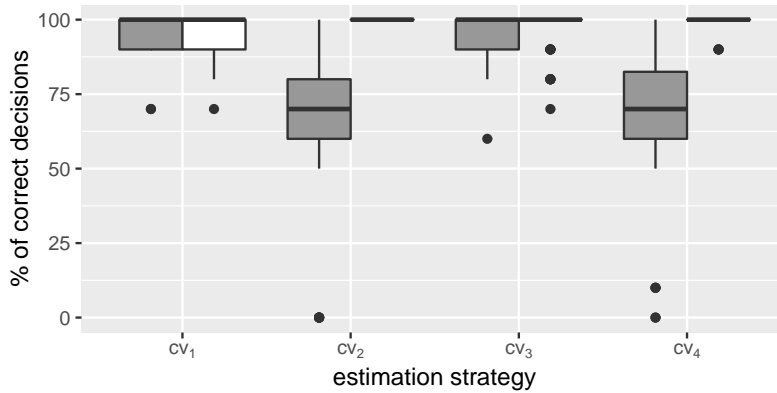}
 \caption{Comparison of different strategies for the estimation of the critical value $cv$. For each $cv$ two boxplots are displayed. The left (gray) boxplot represents the results when two clusters are sampled from two different groups and the right (white) boxplot when the two clusters are sampled from the same group}
    \label{fig:CriticalValue}
  \end{center}	
\end{figure}

The results of this experiment are presented in Figure~\ref{fig:CriticalValue}. For all strategies, the percentage of correct decisions is higher when the two investigated clusters are sampled from the same group, see the right (white) boxplot for each strategy, in comparison to the situation when the clusters are sampled from different groups,  see  the left (gray) boxplot for each strategy. In general, the robustly estimated critical values, i.e. $cv_1$ and $cv_3$, outperform their standard counterparts. 
Since there is no clear difference between the two robust strategies, we select $cv_1$ as the estimation of the critical values for all our following experiments.

\subsection{Number of nearest neighbors}\label{subsubsec:exp:nn}

The aim of the second experiment is to determine the optimal range for the number of nearest neighbors,  $q=1,2,\dots,q_{max}$, considered in the  merging procedure. We investigate three options for the maximal number of nearest neighbors, $q_{max}=5,10,15$. We employ the same clusters as in the previous experiment and the percentage of correct decisions as an indicator for the optimal choice of $q_{max}$.

\begin{figure}[h]
  \begin{center}
    \includegraphics[width=0.65\textwidth]{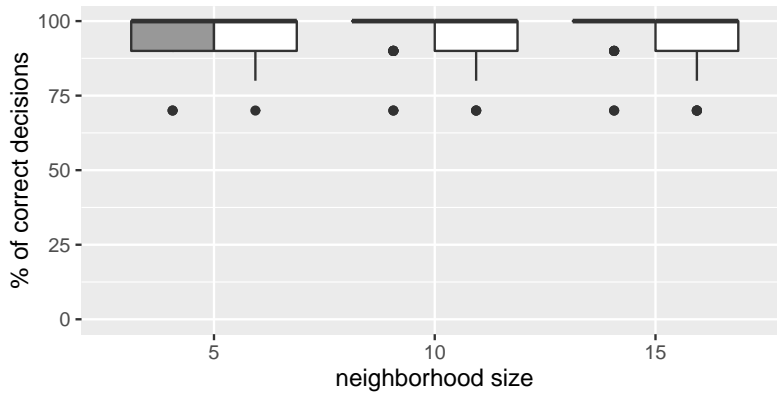}
    \caption{
Comparison of different maximal numbers of nearest neighbors, $q_{max}$. Two boxplots are shown for each $q_{max}$. The left (gray) boxplot represents the results when two clusters are sampled from two different groups and the right (white) boxplot when the two clusters are sampled from the same group}
    \label{fig:neighbors}
  \end{center}
\end{figure}

Figure \ref{fig:neighbors} summarizes the results of the experiment. In general, all choices of $q_{max}$ lead to a high percentage of correct decisions and 
no clear difference can be observed. As a result, we choose $q_{max}=5$ for all our following experiments for computational reasons.

\subsection{Number of initial clusters}\label{NumberCl}

The goal of this experiment is to identify the best strategy for the selection of the number of initial clusters, $k_{init}$, employed in the first step of the proposed IClust algorithm. The parameter is used to partition a given data set into a number of potentially highly pure clusters which are successively merged in the second step of IClust.
In general, $k_{init}$ is supposed to be larger than the actual number of groups in the data set. The actual number of groups is usually not known in advance and the only available information about the data set is the sample size, $n$. Therefore, we determine $k_{init}$ as a function of $n$. One possible approach is to set $k_{init}$ to be  linear dependent on $n$, e.g. $k_{init}=n/4$. However, if the size of a data set is very large, the parameter $k_{init}$ will get considerably high which will notably increase the computational effort of the merging procedure. 
In addition, a large $k_{init}$ value leads to a small size of 
the initial clusters which might affect the efficiency of the merging procedure.

We investigate various options for $k_{init}$ to be non-linearly dependent on~$n$, $k_{init}=  5 \log(n), 10 \log(n),15 \log(n)$. This experiment is again based on $10$ replications of an imbalanced data set sampled from audio data. In each replication, we randomly select $7$ audio groups with $3$ bigger groups of the size:~$100,75,50$, and $4$ smaller groups of the size: $4,3,2,1$, resulting in $235$ observations  in total. We used  Ward's hierarchical clustering algorithm to obtain the initial set of clusters. 

In order to assess the influence of the different settings for $k_{init}$ on the final clustering solution, 
we select several well-known evaluation measures: Purity~\citep{Zhao02criterionfunctions}, F-measure~\citep{DBLP:conf/kdd/LarsenA99} combining the concepts of precision and recall, and V-measure \citep{DBLP:conf/emnlp/RosenbergH07} incorporating homogeneity  and completeness scores. Such measures evaluate a clustering solution as a whole and do not reflect the ability of a clustering method to correctly detect small groups. For this reason, we also employ weighted measures which can assess the performance of a clustering method in terms of detecting small and big groups separately. Table \ref{tab:measures} contains all employed measures, more details about the measures are provided in the 
supplementary material. All measures range between zero and one with higher values indicating a good clustering result and lower values corresponding to a poor clustering solution.  Additionally, we provide two reference values representing potential extremes. The first value corresponds to a clustering solution when all detected clusters are of size one, while the second value corresponds to a clustering solution when all observations are assigned to a single cluster. We report the results before and after applying the merging procedure to demonstrate the performance of the IClust approach in comparison to the initial clustering solution.

\begin{table}[h]
\caption{Overview of employed evaluation measures with corresponding abbreviations~(abbr)}
\centering
\begin{tabular}{l l l l }
\hline
evaluation measure &  abbr 	& weighted evaluation measure & abbr \\
\hline
purity	& $P$ 	& F-measure - big groups & $wF^b$\\
 F-measure & $F$ & precision - big groups & $wPr^b$\\
 V-measure & $V$ &  recall - big groups & $wRe^b$\\
 homogeneity & $H$ & F-measure - small groups & $wF^s$\\
 completeness & $C$ & precision - small groups & $wPr^s$\\
&& recall - small groups & $wRe^s$\\
 \hline
\end{tabular}
\label{tab:measures}
\end{table}

\begin{figure}[h]
  \begin{center}
    \includegraphics[width=1\textwidth]{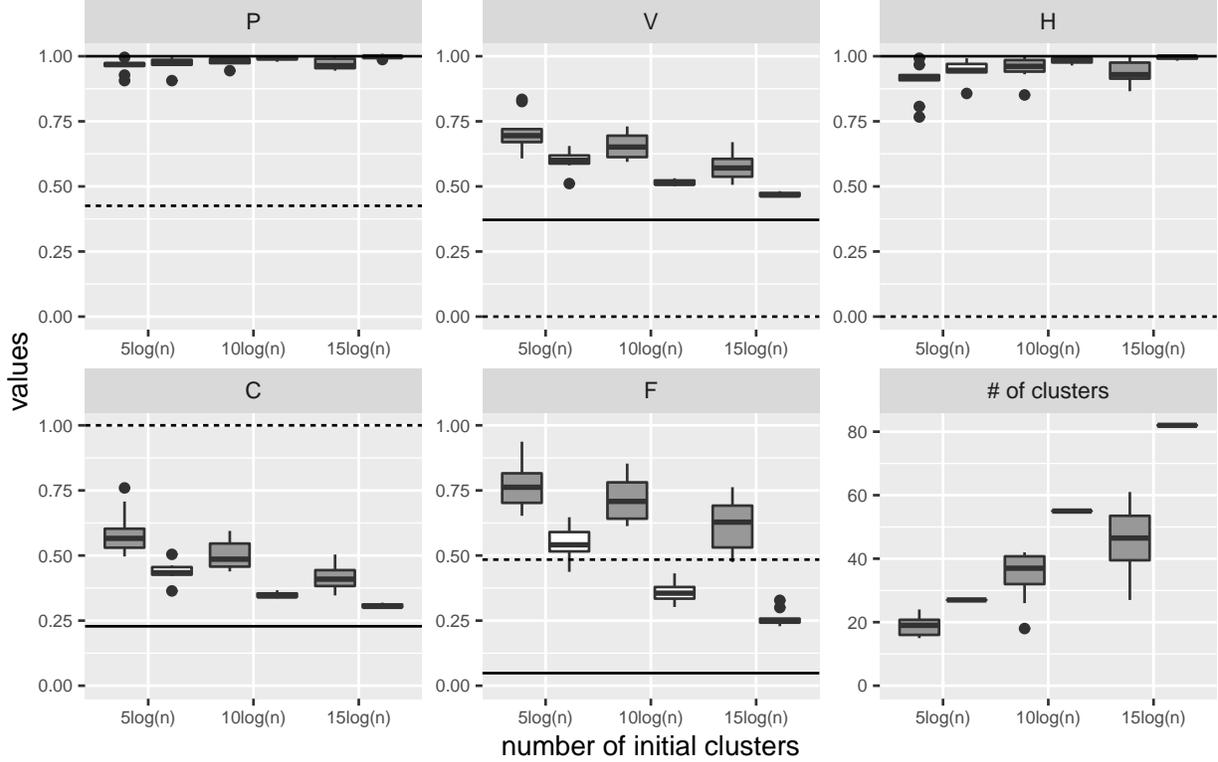}
    \caption{
Comparison of the clustering results for different numbers of initial clusters, $k_{init}$, using purity ($P$), V-measure ($V$), homogeneity ($H$), completeness ($C$), F-measure ($F$), and number  (\#) of clusters. For each $k_{init}$ two boxplots are displayed. The results of the final clusterings correspond to the left (gray) boxplot and the results of the initial clusterings to the right (white) boxplot. The lines indicate two extreme clustering solutions. Solid lines: all clusters are of size one. Dashed lines: all observations are assigned to a single cluster}
    \label{fig:NumberOfInitialClusters}
  \end{center}
\end{figure}

Figure~\ref{fig:NumberOfInitialClusters} depicts the results of the experiments using the conventional clustering evaluation measures. In general, IClust always improves the initial clustering solution indicated by the notable raise of the F-measure ($F$) and the V-measure ($V$) which is directly influenced by the completeness ($C$) of the corresponding clustering solution. Additionally, the scores for $P$ and $H$ show that the purity and homogeneity of the initial clusters are comparable to those of the final clusters. In general, a low number of initial clusters, $k_{init}= 5 \log (n)$, leads to a high completeness ($C$) but also to a lower homogeneity ($H$) of the final clusters, i.e. final clusters partly consists of observations from different groups.  On the opposite, a higher number of initial clusters, $k_{init}= 15 \log(n)$, results in purer clusters (see $P$ and $H$). Additionally, $V$, $C$, and $F$ are only slightly lower than for $k_{init}= 10 \log(n)$. This may indicate that $k_{init}=10 \log (n)$ is the proper choice. To further explore the influence of the different settings for $k_{init}$, we additionally consider the weighted clustering evaluation measures. 

\begin{figure}[h]
  \begin{center}
    \includegraphics[width=1\textwidth]{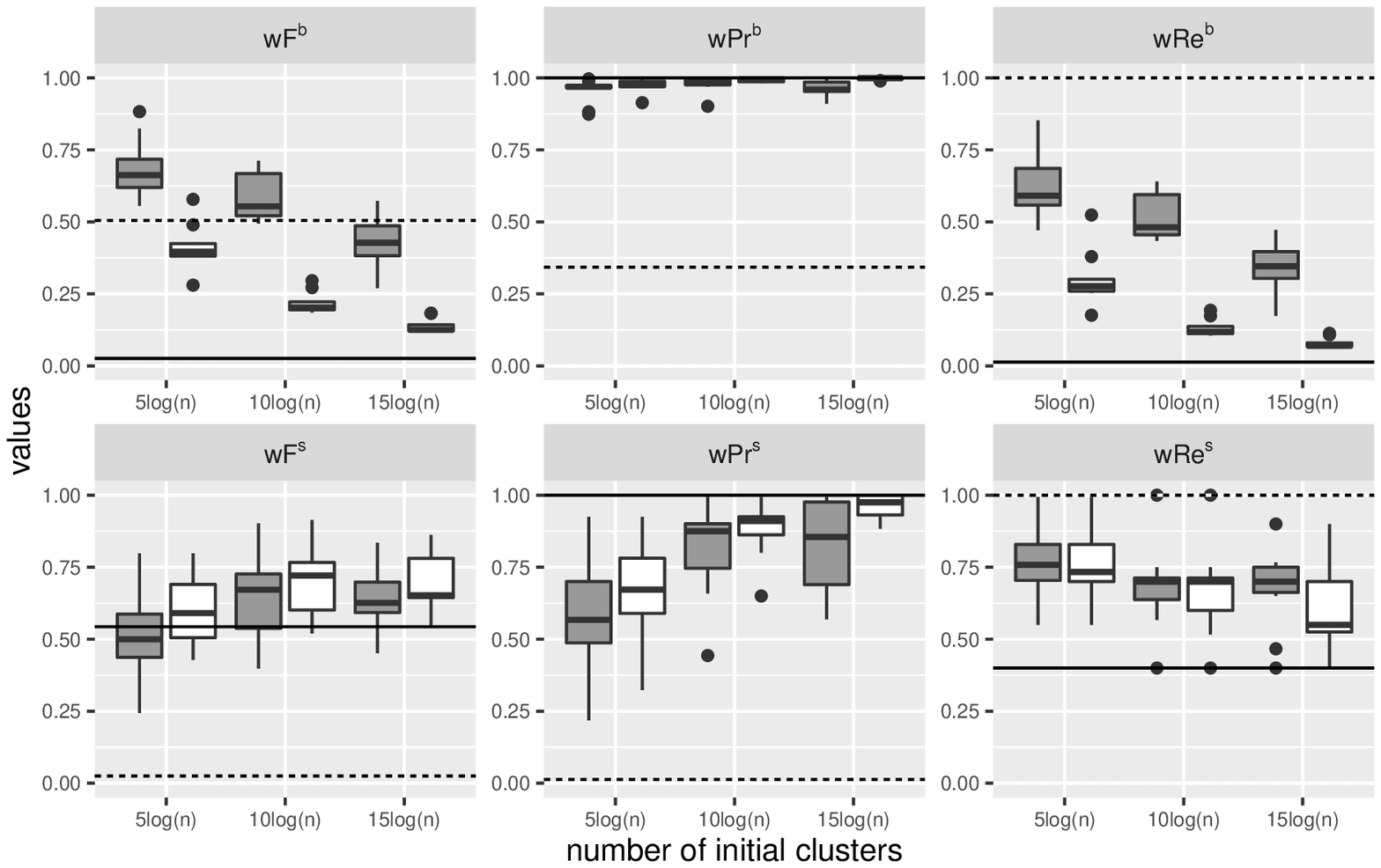}
    \caption{
Comparison of the  clustering solutions for different numbers of initial clusters, $k_{init}$, using the weighted measures for F-measure ($wF$), precision ($wPr$), and recall ($wRe$) with respect to small ($^ s$) and big ($^ b$) clusters.  For each $k_{init}$ two boxplots are displayed. The results of the final clusterings correspond to the left (gray) boxplot and the results of the initial clusterings to the right (white) boxplot. The lines indicate two extreme clustering solutions. Solid lines: all clusters are of size one. Dashed lines: all observations are assigned to a single cluster}
    \label{fig:NumberOfInitialClusters1}
  \end{center}
\end{figure}

Figure~\ref{fig:NumberOfInitialClusters1} summarizes the results of the weighted evaluation measures. The high scores for $wPr^b$  indicate highly pure clusters containing observations from bigger groups independently of $k_{init}$. However, the corresponding recall ($wRe^b$) decreases with an increasing number of initial clusters which reveals that bigger groups are represented by several clusters in the final clustering solution.
This indicates that there are not enough observations in most 
initial clusters leading to difficulties for a proper merging.
As a result, a high number of initial cluster, $k_{init}= 15 \log(n)$, results in a lower F-measure ($wF^ b$) in comparison to the extreme situation all observations build a single final cluster. With respect to small clusters, with an increasing number of initial clusters the precision, $wPr^ s$, increases while the recall, $wRe^s$, decreases slightly. A too low number of initial clusters, such as $k_{init} = 5 \log(n)$, results in a poor clustering solution indicated by the lower median of $wF^s$ in comparison to the potential extreme situation with all final clusters of size one. Therefore, we set $k_{init}=10 \log(n)$ for all our following experiments.

\subsection{Initial clustering algorithm}\label{subsubsec:exp:clusteringMethods}

The last experiment focuses on the selection of the starting clustering algorithm employed in the first step of IClust to partition the provided data set into a number of initial clusters. For this evaluation we consider the following clustering methods (see Section~\ref{subsub:init_method}): $k$-means (KM), Partitioning Around Medoids (PAM), Mclust (MC), and a hierarchical clustering with Ward's criterion (W) and complete linkage (CL). The experiment is again based on $10$ replications of an imbalanced data set randomly sampled from the audio data. The employed data sets are identical to the data sets in the previous experiment. 

\begin{figure}[h]
  \begin{center}
    \includegraphics[width=0.95\textwidth]{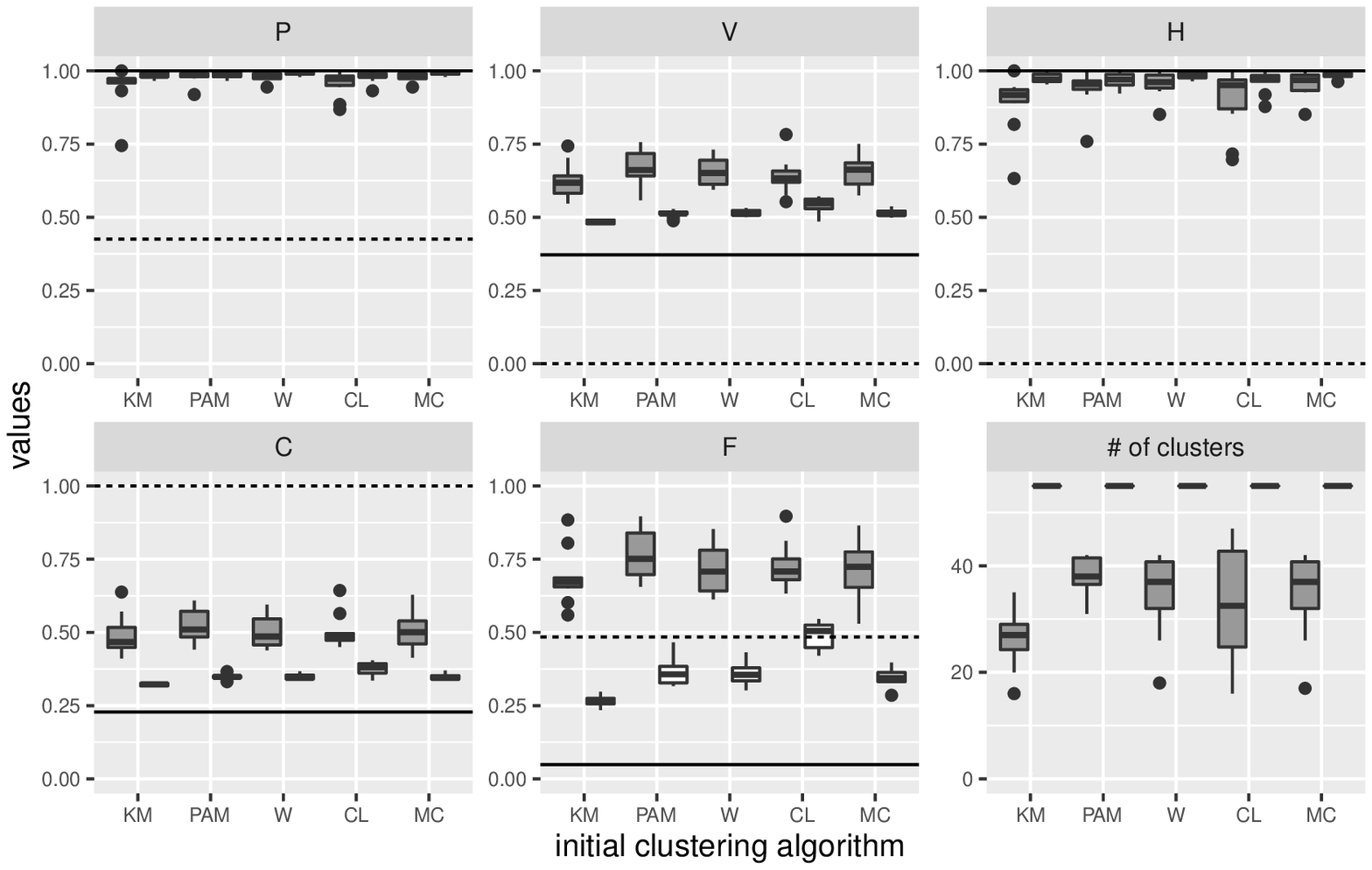}
  \caption{
Comparison of the performance of different initial clustering algorithms, $k$-means (KM), PAM, Mclust (MC), and  a hierarchical clustering with Ward's criterion (W) and complete linkage (CL) using the conventional clustering evaluation measures: purity ($P$), V-measure ($V$), homogeneity ($H$), completeness ($C$), F-measure ($F$), and number (\#) of clusters. Two boxplots are shown for each algorithm. The results of the final clusterings correspond to the left (gray) boxplot  and the results of the initial clusterings to the right (white) boxplot. The lines indicate two extreme clustering solutions. Solid lines: all clusters are of size one. Dashed lines: all observations are assigned to a single cluster}
   \label{fig:Method}
 \end{center}
\end{figure}

Figure~\ref{fig:Method} presents the results using the conventional clustering evaluation measures. In general, the proposed IClust algorithm improves the initial clustering solution independently of the employed starting clustering method. 
This seems to confirm our assumption that IClust can enhance the performance of methods suffering from the uniform effect.
The high scores for purity ($P$) and homogeneity ($H$) indicate the IClust results in highly pure clusters. Although the $F$ measure indicates slightly better clustering results for PAM than for its counterparts, overall, the results do not show any clear differences across the employed starting clustering algorithms.

\begin{figure}[h]
  \begin{center}
    \includegraphics[width=1\textwidth]{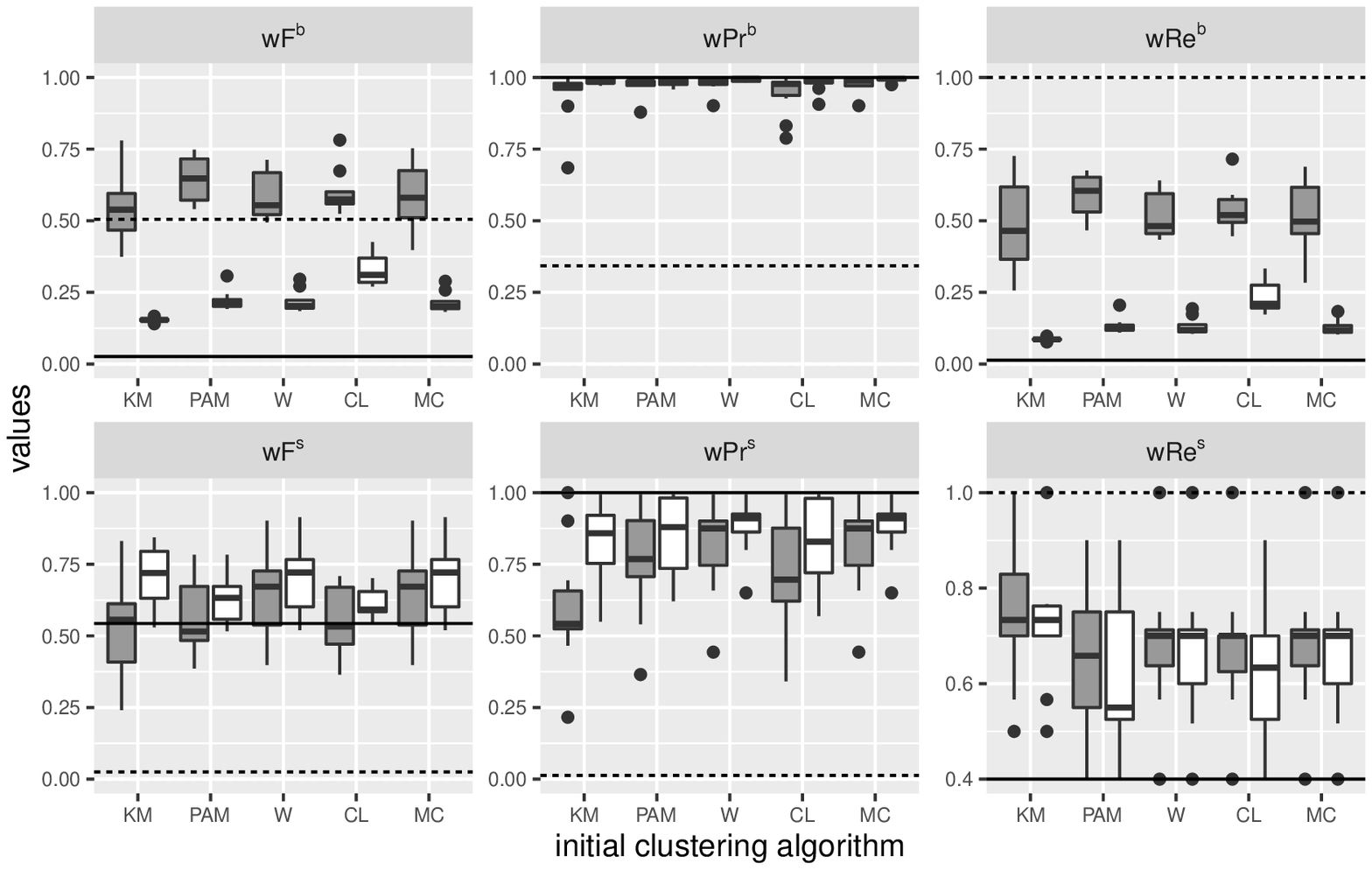}
\caption{
Comparison of the performance of different initial clustering algorithms, $k$-means (KM), PAM, Mclust (MC), and  a hierarchical clustering with Ward's criterion (W) and complete linkage (CL) respectively using the weighted measures for F-measure ($wF$), precision ($wPr$), and recall ($wRe$) with respect to small ($^s$) and big ($^b$) clusters. Two boxplots are shown for each algorithm. The results of the final clusterings correspond to he left (gray) boxplot and the results of the initial clusterings to the right (white) boxplot. The lines indicate two extreme clustering solutions. Solid lines: all clusters are of size one. Dashed lines: all observations are assigned to a single cluster}
    \label{fig:Method1}
  \end{center}
\end{figure}

Figure \ref{fig:Method1} summarizes the results using the weighted clustering evaluation measures. The high precision scores ($wPr^ b$) indicate that clusters containing observations from bigger groups are highly pure. However, the lower recall ($wRe^ b$) and, in following, the lower F-measure ($wF^ b$) show that bigger groups are commonly partitioned into multiple clusters. The PAM clustering method slightly outperforms the other employed methods in terms of $wF^ b$. However, $wF ^s$ indicates that PAM cannot reveal smaller groups since the median is below the value representing the extreme clustering result with all final clusters of size one. Similarly, complete linkage (CL) and $k$-means (KM) achieve a~$wF^ s$ close to the extreme clustering solution. Both methods generate clusters containing observations from more than a single group. Therefore, the precision with respect to smaller groups ($wPr^ s$) achieves low scores and directly influences the F-measure ($wF^ s$). Ward's method (W) and Mclust (M)
better facilitate the detection of small groups in terms of $wF^ s$. For all our following experiments we select  Ward's  method (W) as the starting clustering approach since it is less computationally demanding than the Mclust (MC) algorithm.

\section{Experimental comparison}\label{sec:comparison}

In this section, we compare the proposed IClust approach to several clustering methods on four real-media data sets. In our comparison, we considered several existing, well-established clustering methods: Affinity Propagation (AP)\footnote{The R implementation is available in \texttt{apcluster} \citep{apcluster}}~\citep{Frey07clusteringby}, Mclust (MC)\footnote{The R code is available in \texttt{mclust} \cite{mclust}}~\citep{Fraley00model-basedclustering}, and  $x$-means(XM)\footnote{The R implementation is available at~\url{http://www.rd.dnc.ac.jp/~tunenori/src/xmeans.prog}.}~\citep{DBLP:conf/ideal/Ishioka00}. We restrict the selection of compared methods to parameter-free approaches to ensure that a clustering solution is not affected by a wrong choice of parameters since a parameter setting commonly depends on the nature of the underlying data. The empirical simulation study in Section~\ref{sec:parameters} indicated that the Ward's algorithm (W) seems to be an optimal initial clustering approach. Therefore, we also include the method in our final comparison.  The number of clusters for the W method is taken as the true number of groups (WT) and it is also estimated using three clustering indices\footnote{All clustering indices are implemented in the R package \texttt{clusterSim}~\citep{clusterSim}.}: \emph{Davies-Bouldin} (WDB), \emph{gap statistic} (WG), and \emph{Silhouette}~(WS)~\citep{ROUSSEEUW198753}. In order to have a fair comparison, the upper boundary of a predefined range of the number of clusters is set to the same number of initial clusters employed in IClust, i.e. $10\log(n)$.  Note that the same upper boundary is considered for  Mclust (MC) which estimates the optimal number of clusters based on the largest BIC value. IClust is employed with the previously determined settings, $cv=cv_1$, $q=1,2,\dots,5$,  $k_{init}=10\log(n)$, and  Ward's hierarchical clustering as the initial clustering approach.

In addition to audio data, we employ three media data sets publicly available in UCI machine learning repository\footnote{UCI Machine Learning Repository: \url{http://archive.ics.uci.edu/ml/}}, see Table \ref{tab:realdata}. For each media set we randomly select observations from the original groups to construct a similar imbalanced data sets. The ratios between group sizes are kept the same among the constructed datasets, but the sizes of the groups are different. The idea behind this setup is to see whether or not the compared methods are affected by different amounts of information, i.e. the number of observations, in the groups. All experiments are based on $10$ replications. We report the results using five clustering evaluation measures:  V-measure ($V$), homogeneity ($H$), completeness ($C$), as well as  the weighted F-measures with respect to big and small clusters ($wF^b$ and $wF^ s$). The supplementary material includes the complete evaluation results including the measures omitted in this section.
\begin{table}[h]
  \begin{center}
    \caption{Overview of the employed real-world data sets in terms of number of observations ($n$), dimensionality ($p$), and number (\#) of groups}\label{tab:realdata}
    \begin{tabular}{l crr}
      \hline
      & \multirow{2}{*}{   $n \times p~\times $\#groups}	&   	    \multicolumn{2}{c}{group size}\\
      \cline{3-4}
      &     &$\min$ &$\max$\\
      \hline
      Audio 	&  
      \multirow{4}{*}{
      $\begin{matrix}
      	  4780 & \times & 679 & \times & 12\\
	  10299 & \times & 561 & \times & 6 \\
	10992 & \times &   16 & \times & 10  \\
	  6435 & \times &   36 & \times  & 6  
      \end{matrix}$}
         &  102&2164\\
      Human Activity Recognition  &	 & 1406 & 1944\\
      Pen-Based Recognition	& 	& 1055 &1144 \\
      Statlog Landsat Satellite  	& 	& 626 &1533\\
      \hline
    \end{tabular}
  \end{center}
\end{table}

\subsection{Comparison on the audio data}\label{subsubsec:exp:final:audio}

We first construct 10 imbalanced data sets from the high-dimensional audio data. Each setting includes $10$ groups with $3$ bigger groups of the sizes: $100, 75, 50$, and $7$ smaller groups of the sizes: $4, 3, 3, 2, 2, 1, 1$, resulting in $241$ observations in total. 

\begin{figure}[h]
  \begin{center}
    \includegraphics[width=1\textwidth]{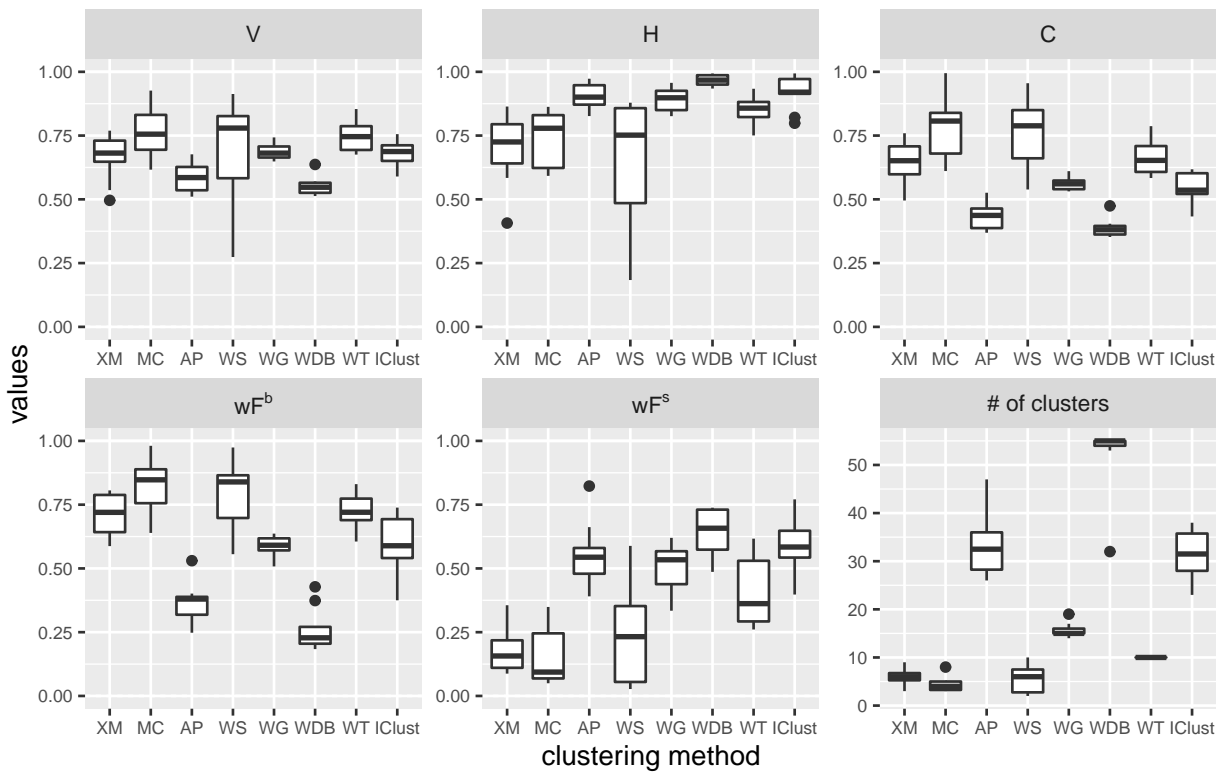}
     \caption{Clustering results on the audio data set in terms of V-measure ($V$), homogeneity ($H$), completeness ($C$), the weighted F-measures $wF^ b$ and $wF^ s$ with respect to big and small groups respectively, and the number (\#) of detected clusters}
     \label{fig:Audio}
   \end{center}
\end{figure}

Figure~\ref{fig:Audio} 
indicates poor performance of some centroid-based approaches, such as $x$-means (XM) and the Ward's method with Silhouette Width (WS), in clustering highly imbalanced media (audio) groups. The methods detect a lower number of clusters than the actual number of groups (cp. WT approach)
leading to low homogeneity~($H$). High values $wF^b$ indicate that the methods can reasonably reveal bigger groups but low $wF^s$ show difficulties regarding the detection of smaller groups. Similarly, the Mclust (MC) cannot reveal small audio groups indicated by the lowest $wF^s$.  In contrast, high $wF^b$
indicate appropriate handling of bigger groups. 
The performance of Ward's method with the true number of groups (WT) seems to be also affected by the presence of strongly varying group sizes.  Even though the method still generates homogeneous clusters (see $H$), slightly low $wF^s$
 indicate difficulties in identifying very small groups.
 Although WDB achieves the highest homogeneity ($H$) and weighted F-measure with respect to smaller groups ($wF^ s$), the underlying clustering solutions are suboptimal due to the high number of final clusters leading to the lowest completeness ($C$) and consequently low V-measure ($V$).
Surprisingly, Ward's method with the Gap statistic (WG) appears to reasonably detect both smaller and larger audio groups (see $wF^s$ and $wF^b$).

The proposed IClust approach outperforms WG in terms of revealing smaller groups (see $wF^s$). In addition, IClust is capable of finding bigger groups as well (see $wF^b$) in comparison to methods achieving high homogeneity ($H$), such as WG and Affinity Propagation (AP).

\subsection{Comparison on the pen-based recognition data}\label{subsubsec:exp:final:pen}

The second experiment employs 10 imbalanced data sets constructed from the pen-based recognition data. Each setting contains $10$ groups with $3$ bigger groups of the sizes: $1000, 750, 500$, and $7$ smaller groups of the sizes: $40, 30, 30, 20, 20, 10, 10$, resulting in a total sample size of $2380$.

\begin{figure}[h]
  \begin{center}
    \includegraphics[width=1\textwidth]{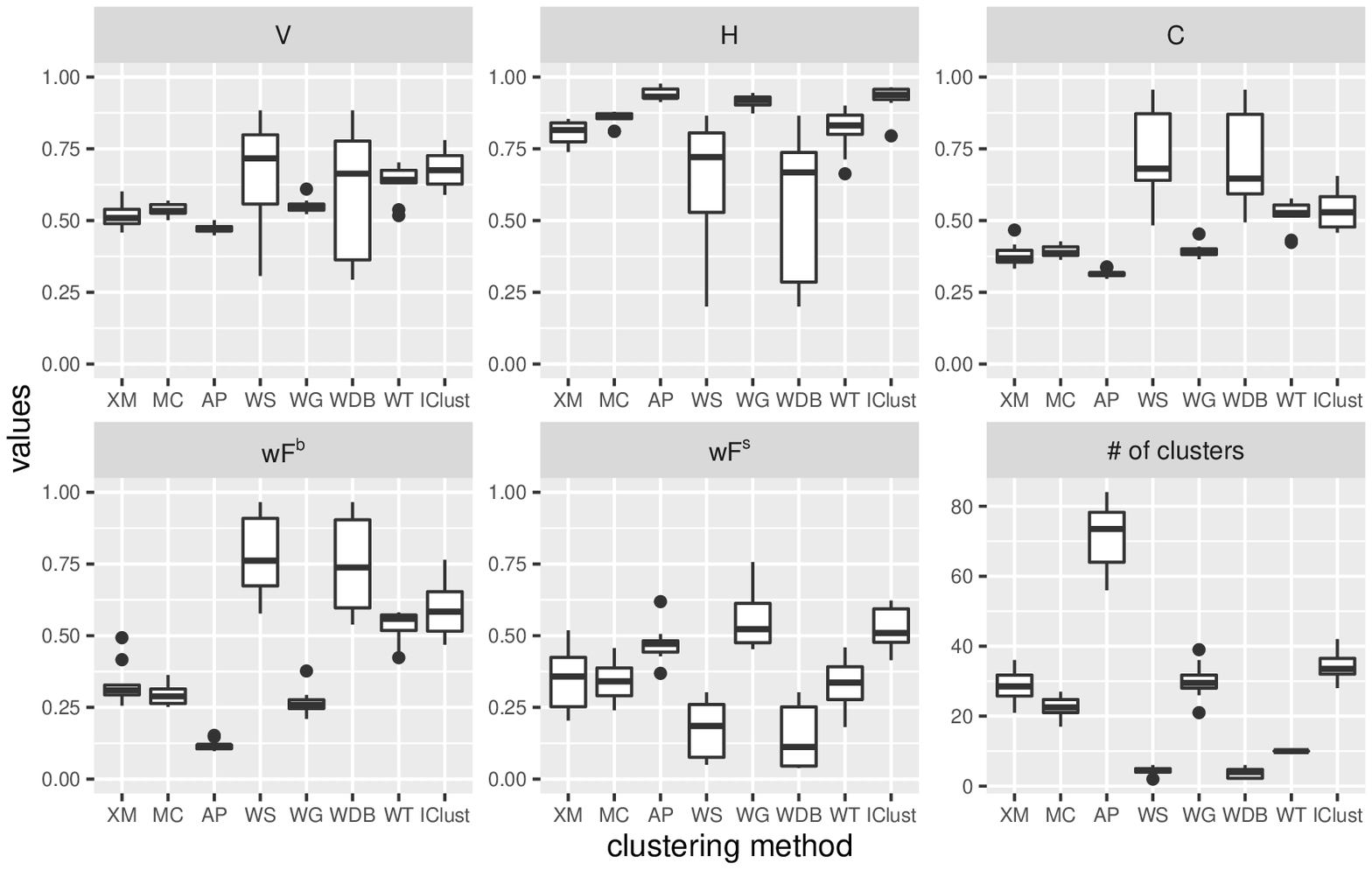}
      \caption{Clustering results on the pen-based recognition data set in terms of V-measure ($V$), homogeneity ($H$), completeness ($C$), the weighted F-measures $wF^ b$ and $wF^ s$ with respect to big and small groups respectively, and the number (\#) of detected clusters}
    \label{fig:PenDigits}
  \end{center}
\end{figure}

Figure~\ref{fig:PenDigits} shows that Ward's method with the Silhouette Width (WS) and the Davis-Bouldin index (WDB) seem to have troubles to reveal small groups
indicated by low $wF^s$. Moreover, the methods produced a considerably lower number of clusters leading to low homogeneity ($H$) scores. Although the model-based MC and $x$-means (XM) generate to some extent homogeneous clusters (see $H$), low $wF^b$ as well as low $wF^s$ demonstrate that the methods completely fail in detecting the considered media groups.

As expected, the proposed IClust method appears to identify both smaller and bigger groups indicated by high $wF^s$ and $wF^b$.  Although IClust produces a larger number of clusters than the true number of groups leading to low completeness ($C$), the methods outperform the remaining approach (i.e.~WT, AP and WG) in terms of the V-measure ($V$). 

\subsection{Comparison on the human activity recognition data}\label{subsubsec:exp:final:activity}

The next comparison is applied on 10 imbalanced data sets randomly sampled from the human activity recognition data. Each setting consists of $6$ groups with $3$ bigger groups of the sizes: $200, 150, 100$, and $3$ smaller groups of the sizes: $8, 6, 4$, resulting in a total sample size of $468$. 

\begin{figure}[h]
  \begin{center}
    \includegraphics[width=1\textwidth]{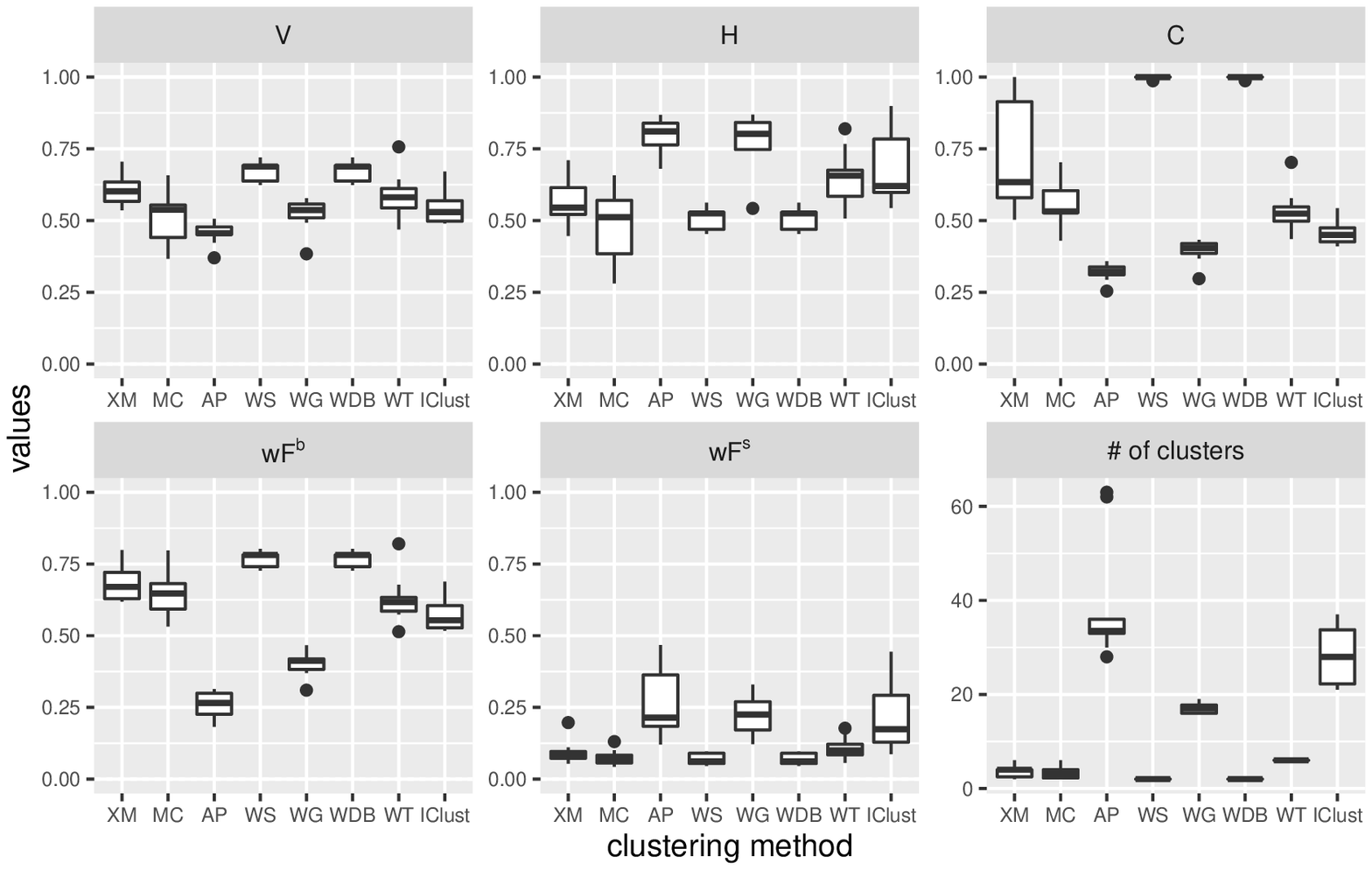}
    \caption{Clustering results on the human activity recognition data set in terms of V-measure ($V$), homogeneity ($H$), completeness ($C$), the weighted measures $wF^ b$ and $wF^ s$ with respect to big and small groups respectively, and the number (\#) of detected clusters}
    \label{fig:Human}
  \end{center}
\end{figure}

Figure \ref{fig:Human} shows that Ward's method with Silhouette Width (WS) and Davies-Bouldin (WDB) index have again difficulties in clustering imbalanced media groups as in the previous experiment.  Similarly, the performance of $x$-means (XM) and the model-based MC appears to be violated by strongly varying group sizes indicated by low homogeneity ($H$). 
The performance of Ward's method (WT) seems to be also affected like in case of audio data indicated by low $wF^s$.

The proposed IClust algorithm is slightly worse than the best performing methods, such as AP and WG, in terms of generating homogeneous clusters (see $H$). The methods also detect more clusters than the true number of groups (cp. WT approach) leading to low completeness ($C$) and the V-measure ($V$). 
All three methods demonstrate the best performance regarding the detection of small groups (see $wF ^s$). 

\subsection{Comparison on the satellite data}\label{subsubsec:exp:final:satellite}

The last experiment compares the employed methods on 10 imbalanced data sets randomly generated from the satellite data. For each setting we sample $6$~groups with $3$ bigger groups of the sizes: $300, 225, 150$, and $3$~smaller groups of the sizes: $12 , 9, 3$, resulting in $669$ observations in total. 

\begin{figure}[h]
  \begin{center}
    \includegraphics[width=1\textwidth]{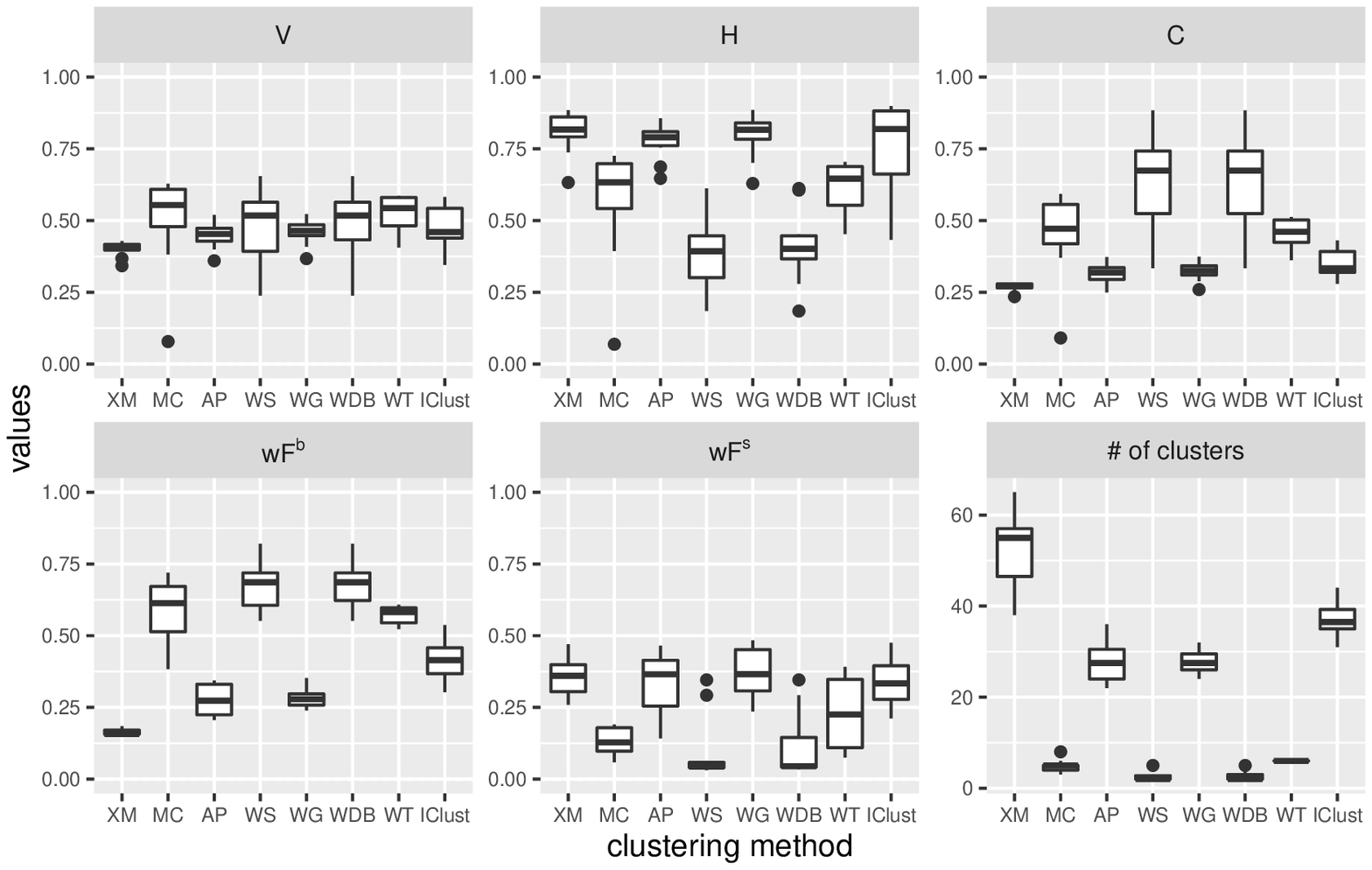}
    \caption{Clustering results on the satellite data set in terms of V-measure ($V$), homogeneity ($H$), completeness ($C$), the weighted measures $wF^ b$ and $wF^ s$ with respect to big and small groups respectively, and the number (\#) of detected clusters}
    \label{fig:Satellite}
  \end{center}
\end{figure}

Figure~\ref{fig:Satellite} shows that identifying media groups is again challenging for WS, WDB and MC indicated by low homogeneity ($H$). The detection of small groups is more problematic than revealing larger groups (see $wF^s$ versus $wF^b$) for these methods. Surprisingly,  $x$-means (XM) generates highly homogeneous clusters. However, this is caused by
over-clustering the considered media data sets  
leading to the lowest completeness ($C$) and thus low V-measure ($V$).
The low homogeneity ($H$) scores for WT demonstrate that a prior knowledge about the actual number of groups does not necessarily lead to a correct clustering solution in a highly imbalanced scenario.

The proposed IClust algorithm outperforms Affinity Propagation (AP) and the Ward's method with the Gap statistic (WG) with respect to handling bigger clusters in terms of $wF^b$. Regarding the detection of smaller groups, IClust demonstrates comparable performance (see $wF^s$).

\section{Discussion and conclusions}\label{sec:discussion_conclusion}

We summarize our main findings based on the quality of the compared methods in terms of the weighted F-measures ($wF^b$ and $wF^s$).  Unlike conventional clustering evaluation measures, the weighted F-measures allow  to inspect the ability of clustering methods to identify big and small groups. 

\begin{figure}[h]
  \begin{center}
    \includegraphics[width=0.8\textwidth]{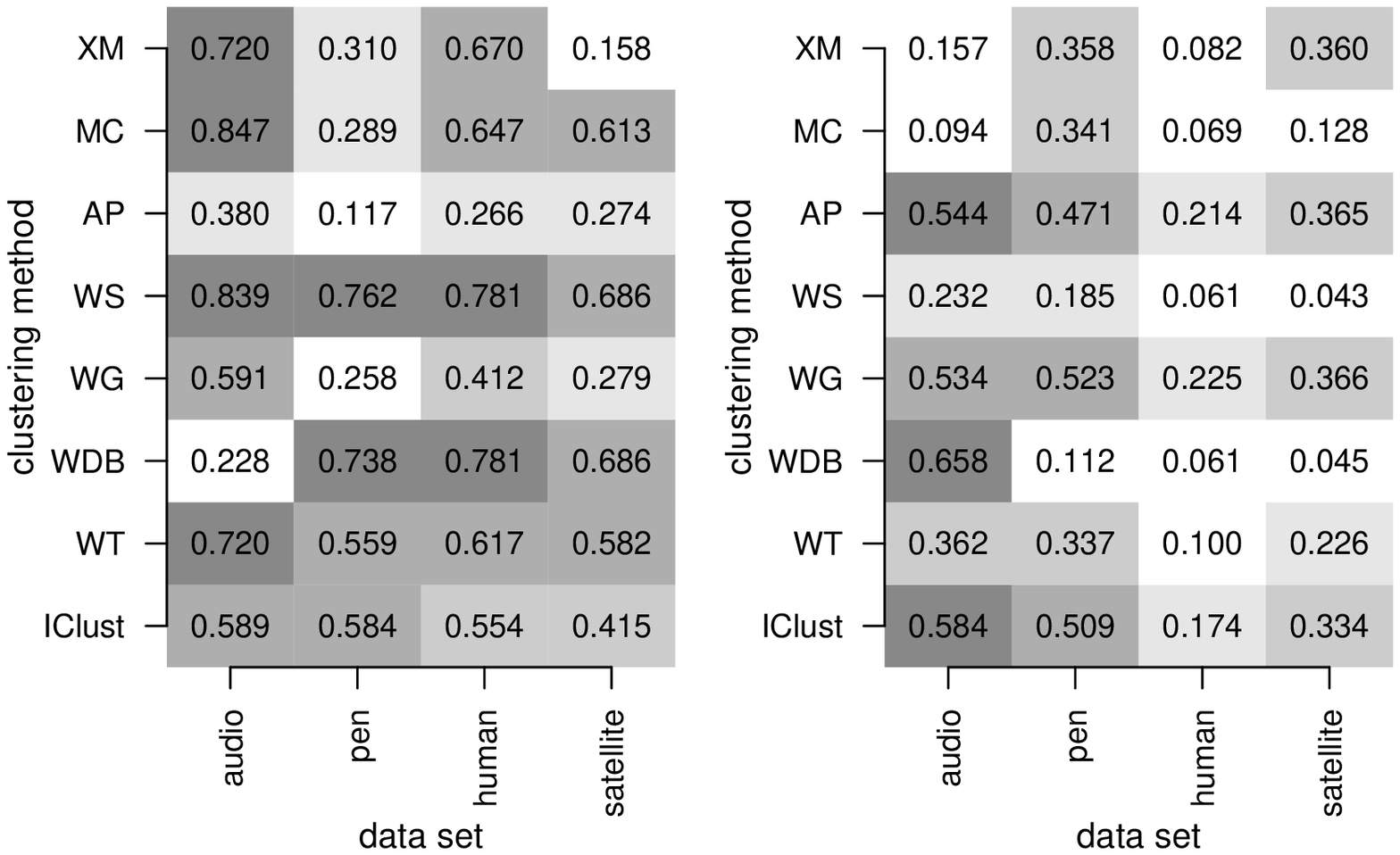}
    \caption{Summary of all employed experiments in terms of the weighted F-measures~$wF^b$ (left) and $wF^s$ (right) with respect to big and small groups. Dark shades of gray indicate good performance and light shades of gray indicate poor performance }\label{fig:summary}
  \end{center}
\end{figure}

Figure~\ref{fig:summary} shows the obtained results in terms of the median performance of $wF^b$ and $wF^s$ achieved during the previously performed experiments.  Overall, there is no clear dependence between the performance of the methods and the sizes of employed data sets. However, the results indicate  a dependency between identifying bigger and smaller groups. The methods capable of identifying big groups, such as WS and WDB, show considerably weaker ability to detect small groups. 
This supports the fact that centroid-based methods as well as validity indexes
might not be suitable for clustering imbalanced high-dimensional data. Similarly, the model-based MC demonstrates poor performance in terms of finding smaller groups.
On the opposite, Affinity Propagation (AP), which does not assume any specific cluster characteristics, appears to be among the best performing methods in terms of revealing small groups (see dark shades of gray in Figure \ref{fig:summary}, right). However, the method turns out to poorly detect bigger groups (see light shades of gray in Figure \ref{fig:summary}, left).  

In contrast to all investigated method, the proposed IClust approach shows the best performance in terms of finding small groups and, in addition,  the method can still reasonably identify bigger groups (see dark shades of gray in Figure \ref{fig:summary}, left). 

Although any real-world data set exhibits a multiple group structure, it may be difficult to determine, whether or not the groups are of different sizes if there is no prior knowledge available. This leads to the question if the proposed IClust algorithm can identify groups of approximately the same sizes (balanced data setting). For this reason, we perform an additional experiment on the balanced pen-based recognition data set which was used to evaluate various clustering methods for high-dimensional data \citep{DBLP:journals/pvldb/MullerGAS09}. \citet{DBLP:journals/pvldb/MullerGAS09} considered 10 groups of  similar sizes resulting in a total size of 7494. Table \ref{tab:balanced} presents the performance of the compared methods in terms of the conventional evaluation measures.  
Although IClust shows slightly worse performance than the best performing XM, MC, AP, and WG in terms of purity ($P$) and homogeneity ($H$), the proposed method partitions the data set into a lower number of clusters. 
This is indicated by higher completeness ($C$) as well as higher V-measure ($V$). In addition, the F-measure suggests that IClust may also be used to reveal a~group structure in a balanced data setting.
\begin{table}[h]
\centering
\caption{Clustering quality of the compared methods on the balanced pen-based recognition data set with respect to purity ($P$), V-measure ($V$), homogeneity ($H$), completeness ($C$), F-measure ($F$), and the number (\#) of clusters}\label{tab:balanced}
\begin{tabular}{ l c c c c c c }
\hline
 & $P$ &  $V$& $H$& $C$ & $F$ & \# of clusters \\
  \hline
                 
XM    &        0.916 &0.653 &0.881 &0.518& 0.496  & 79\\
MC          &  0.937& 0.670 &0.930& 0.523 &0.428  & 72\\
AP          &  0.981& 0.615 &0.971& 0.450& 0.218& 165\\
WS   &0.795& 0.767& 0.774& 0.760& 0.757&  12\\
WG &0.970 &0.700& 0.959 &0.551 &0.422  & 67\\
WDB& 0.626& 0.693 &0.635 &0.763& 0.663 &  8\\
WT  &0.750 &0.759 &0.739 &0.780& 0.736 &  10\\
IClust       &0.823& 0.769& 0.862& 0.693 &0.673& 35\\
\hline
\end{tabular}
\end{table}

The proposed method also has limitations. First, IClust takes four input parameters. Although we provided a thorough empirical study to select optimal parameters, the parameter setting may be tuned for other data sets. Second, IClust tends to generate a larger number of clusters than the actual number of groups in the data. This might be 
due to the estimation of critical values employed in the merging procedure. A possible solution for improvement could be either to adjust the critical values to the size of the clusters which are merged during the procedure or to incorporate different robust counterparts to arithmetic mean and standard deviation.
Despite the mentioned limitations, the proposed IClust
algorithm exhibits also some advantages over existing methods. 
IClust does not require a pre-specification of the number of final clusters. 
This algorithm also does not assume any specific cluster and data characteristics. 
Moreover, the experiments demonstrated that the choice of parameters seems to be 
reasonable in both imbalanced and balanced scenarios. This indicates that
IClust is a useful clustering method for media data, and it is a promising 
method also for other application domains.  The R implementation of the algorithm is also  freely available at \texttt{https://github.com/brodsa/IClust}.

\acknowledgments 
This work has been partly funded by the Vienna Science and Technology Fund (WWTF) through project ICT12-010 and by the K-project DEXHELPP through COMET - Competence Centers for Excellent Technologies, supported by BMVIT, BMWFW and the province Vienna. The COMET program is administrated by FFG.




\begin{thebibliography}{27}
\providecommand{\natexlab}[1]{#1}
\providecommand{\url}[1]{{#1}}
\providecommand{\urlprefix}{URL }
\expandafter\ifx\csname urlstyle\endcsname\relax
  \providecommand{\doi}[1]{DOI~\discretionary{}{}{}#1}\else
  \providecommand{\doi}{DOI~\discretionary{}{}{}\begingroup
  \urlstyle{rm}\Url}\fi
\providecommand{\eprint}[2][]{\url{#2}}

\bibitem[{Bloisi and Iocchi(2008)}]{DBLP:conf/icvs/BloisiI08}
Bloisi DD, Iocchi L (2008) {R}ek-{M}eans: {A} k-means based clustering
  algorithm. In: International Conference on Computer Vision Systems ({ICVS}),
  pp 109--118

\bibitem[{Breunig et~al(2000)Breunig, Kriegel, Ng, and
  Sander}]{DBLP:conf/sigmod/BreunigKNS00}
Breunig MM, Kriegel H, Ng RT, Sander J (2000) {LOF:} identifying density-based
  local outliers. In: {ACM} {SIGMOD} International Conference on Management of
  Data ({ICMD}), pp 93--104

\bibitem[{Fraley and Raftery(2000)}]{Fraley00model-basedclustering}
Fraley C, Raftery AE (2000) Model-based clustering, discriminant analysis, and
  density estimation. Journal of the American Statistical Association
  97:611--631

\bibitem[{Frey and Dueck(2007)}]{Frey07clusteringby}
Frey BJ, Dueck D (2007) Clustering by passing messages between data points.
  Science 315(5814):972--976

\bibitem[{Hartigan and Wong(1979)}]{Hartigan/Wong:79}
Hartigan JA, Wong MA (1979) A {K}-means clustering algorithm. Applied
  Statistics 28:100--108

\bibitem[{Hasan et~al(2009)Hasan, Chaoji, Salem, and Zaki}]{Hasan2009994}
Hasan MA, Chaoji V, Salem S, Zaki MJ (2009) Robust partitional clustering by
  outlier and density insensitive seeding. Pattern Recognition Letters
  30(11):994 -- 1002

\bibitem[{Ishioka(2000)}]{DBLP:conf/ideal/Ishioka00}
Ishioka T (2000) Extended k-means with an efficient estimation of the number of
  clusters. In: International Conference on Intelligent Data Engineering and
  Automated Learning ({IDEAL}). Data Mining, Financial Engineering, and
  Intelligent Agents, pp 17--22

\bibitem[{Kaufman and Rousseeuw(1990)}]{DBLP:books/wi/KaufmanR90}
Kaufman L, Rousseeuw PJ (1990) Finding Groups in Data: An Introduction to
  Cluster Analysis. John Wiley, New York

\bibitem[{Krawczyk(2016)}]{Krawczyk2016}
Krawczyk B (2016) Learning from imbalanced data: open challenges and future
  directions. Progress in Artificial Intelligence 21(9):1--12

\bibitem[{Kriegel et~al(2009)Kriegel, Kr{\"{o}}ger, and
  Zimek}]{DBLP:journals/tkdd/KriegelKZ09}
Kriegel H, Kr{\"{o}}ger P, Zimek A (2009) Clustering high-dimensional data: {A}
  survey on subspace clustering, pattern-based clustering, and correlation
  clustering. ACM Transactions on Knowledge Discovery from Data 3(1):1--58

\bibitem[{Kriegel et~al(2011)Kriegel, Kr{\"{o}}ger, Sander, and
  Zimek}]{DBLP:journals/widm/KriegelKSZ11}
Kriegel H, Kr{\"{o}}ger P, Sander J, Zimek A (2011) Density-based clustering.
  {WIRE}s Data Mining and Knowledge Discovery 1(3):231--240

\bibitem[{Larsen and Aone(1999)}]{DBLP:conf/kdd/LarsenA99}
Larsen B, Aone C (1999) Fast and effective text mining using linear-time
  document clustering. In: {ACM} {SIGKDD} International Conference on Knowledge
  Discovery and Data Mining ({KDD}), pp 16--22

\bibitem[{Maechler et~al(2015)Maechler, Rousseeuw, Struyf, Hubert, and
  Hornik}]{cluster}
Maechler M, Rousseeuw P, Struyf A, Hubert M, Hornik K (2015) cluster: Cluster
  Analysis Basics and Extensions.
  \urlprefix\url{https://cran.r-project.org/web/packages/cluster/}, r package
  version 2.0.3

\bibitem[{M{\"{u}}ller et~al(2009)M{\"{u}}ller, G{\"{u}}nnemann, Assent, and
  Seidl}]{DBLP:journals/pvldb/MullerGAS09}
M{\"{u}}ller E, G{\"{u}}nnemann S, Assent I, Seidl T (2009) Evaluating
  clustering in subspace projections of high dimensional data. Proceedings of
  the {VLDB} Endowment ({PVLDB}) 2(1):1270--1281

\bibitem[{Murtagh and Legendre(2014)}]{Murtagh2014}
Murtagh F, Legendre P (2014) Ward's hierarchical agglomerative clustering
  method: Which algorithms implement ward's criterion? Journal of
  Classification 31(3):274--295

\bibitem[{Parsons et~al(2004)Parsons, Haque, and
  Liu}]{DBLP:journals/sigkdd/ParsonsHL04}
Parsons L, Haque E, Liu H (2004) Subspace clustering for high dimensional data:
  a review. {SIGKDD} Explorations Newsletter 6(1):90--105

\bibitem[{Qian and Saligrama(2014)}]{DBLP:conf/icassp/QianS14}
Qian J, Saligrama V (2014) Spectral clustering with imbalanced data. In: {IEEE}
  International Conference on Acoustics, Speech and Signal Processing
  ({ICASSP}), pp 3057--3061

\bibitem[{{R Core Team}(2016)}]{stats}
{R Core Team} (2016) R: A Language and Environment for Statistical Computing. R
  Foundation for Statistical Computing, Vienna, Austria,
  \urlprefix\url{https://www.R-project.org/}

\bibitem[{Rosenberg and Hirschberg(2007)}]{DBLP:conf/emnlp/RosenbergH07}
Rosenberg A, Hirschberg J (2007) V-measure: {A} conditional entropy-based
  external cluster evaluation measure. In: Joint Conference on Empirical
  Methods in Natural Language Processing and Computational Natural Language
  Learning (EMNLP-CoNLL), pp 410--420

\bibitem[{Rousseeuw(1987)}]{ROUSSEEUW198753}
Rousseeuw PJ (1987) Silhouettes: A graphical aid to the interpretation and
  validation of cluster analysis. Journal of Computational and Applied
  Mathematics 20:53--65

\bibitem[{Walesiak and Dudek(2015)}]{clusterSim}
Walesiak M, Dudek A (2015) clusterSim: Searching for Optimal Clustering
  Procedure for a Data Set.
  \urlprefix\url{https://CRAN.R-project.org/package=clusterSim}, r package
  version 0.44-2

\bibitem[{Wang and Chen(2014)}]{DBLP:conf/icarcv/WangC14}
Wang Y, Chen L (2014) Multi-exemplar based clustering for imbalanced data. In:
  International Conference on Control Automation Robotics {\&} Vision
  ({ICARCV}), pp 1068--1073

\bibitem[{Zhao and Karypis(2002)}]{Zhao02criterionfunctions}
Zhao Y, Karypis G (2002) Criterion functions for document clustering:
  Experiments and analysis. Tech. rep., University of Minnesota

\bibitem[{Zimek et~al(2012)Zimek, Schubert, and Kriegel}]{SAM:SAM11161}
Zimek A, Schubert E, Kriegel HP (2012) A survey on unsupervised outlier
  detection in high-dimensional numerical data. Statistical Analysis and Data
  Mining 5(5):363--387

\end{thebibliography}

\end{document}